\pdfoutput=1  
\ifdefined\pdfminorversion\pdfminorversion=7\fi
\documentclass[conference]{IEEEtran}
\IEEEoverridecommandlockouts  
\usepackage{cite}
\usepackage{amsmath,amssymb,amsfonts}
\usepackage{graphicx}
\usepackage{xcolor}
\usepackage{booktabs}
\usepackage{tabularx}
\usepackage{subcaption}
\newcommand{\SI}[2]{\ensuremath{#1\,\mathrm{#2}}}
\usepackage{url}
\usepackage{adjustbox}
\usepackage{stfloats}    
\fnbelowfloat
\usepackage[hidelinks]{hyperref}
\usepackage{tikz}
\usetikzlibrary{arrows.meta, positioning, fit, calc}
\pgfdeclarelayer{bg}\pgfsetlayers{bg,main}

\newif\ifanonymous
\anonymousfalse

\tikzset{
  blk/.style={rectangle, draw=black, fill=white, rounded corners=1pt, align=center, font=\scriptsize, inner sep=3pt, minimum height=0.75cm},
  stage/.style={rectangle, draw=black!60, dashed, rounded corners=2pt, inner sep=5pt},
  arr/.style={-{Stealth[length=2mm]}, thick},
  lbl/.style={font=\scriptsize\itshape, align=center},
}

\begin{document}

\title{Barrier-Shaped Recurrent Reinforcement Learning for Autonomous Landing on a Heaving Ship Deck 
}

\ifanonymous
\author{\IEEEauthorblockN{Anonymous Authors}
\IEEEauthorblockA{Paper ID 4916}}
\else

\author{%
\IEEEauthorblockN{Ritwik Shankar\textsuperscript{*,1}, Chiranjeev Prachand\textsuperscript{*,2}, Abhishek\textsuperscript{1}, and Soumya Ranjan Sahoo\textsuperscript{2}}
\thanks{\textsuperscript{*}These authors contributed equally.}%
\thanks{The authors are with the \textsuperscript{1}Department of Aerospace Engineering and the \textsuperscript{2}Department of Electrical Engineering, Indian Institute of Technology Kanpur, Kanpur, India. This research was supported by the IndiaAI Fellowship. }%
}
\fi
\maketitle

\begin{figure*}[!b]
    \centering
    \includegraphics[width=\textwidth]{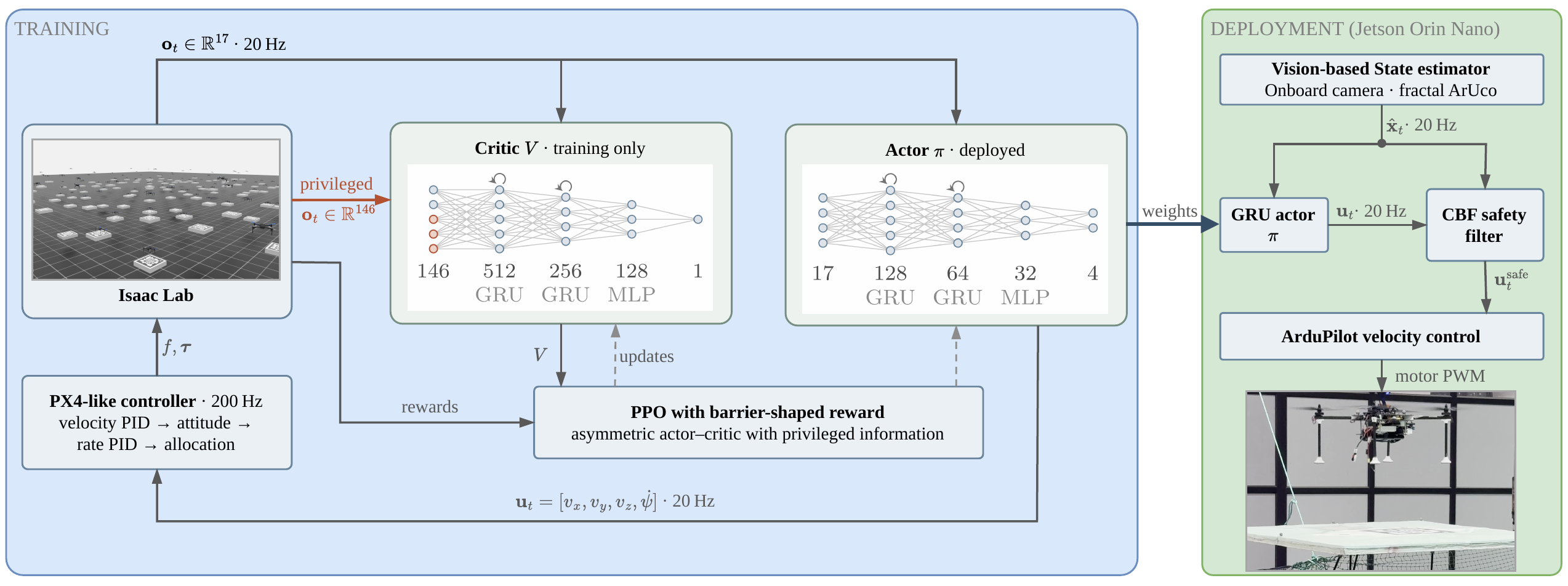}
    \caption{Training and deployment pipeline of the proposed landing algorithm.}
    \label{fig:system}
\end{figure*}
\begin{abstract}
This paper addresses autonomous landing of unmanned aerial vehicle (UAV) rotorcrafts on a heaving ship deck using a recurrent policy trained via asymmetric actor-critic reinforcement learning: the critic sees \SI{6}{s} of future deck height during training, while the actor sees only what the onboard sensors provide in flight, a 17-dimensional state made of the vehicle's position, velocity and attitude relative to the deck and outputs world-frame velocity and yaw-rate commands. Training is performed across 4096 parallel simulated environments, followed by fine-tuning in 16 environments with an onboard vision pipeline in the loop. A control-barrier-function (CBF) stopping margin on the deck-relative vertical state is incorporated at two stages: as a reward term during training, where it halves the median simulated contact speed relative to a policy trained without it, and as a runtime safety filter at deployment, evaluated at every control step to abort and retry the descent when the margin is violated. Because the autopilot's disarm logic cannot detect the vehicle resting on a moving deck, proximity-based thrust cutoff at touchdown is commanded directly in the landing pipeline. The proposed approach is validated using a parallel-manipulator-platform-based deck emulator that reproduces the heaving motion of the ship deck (scaled to \SI{0.70}{m} peak-to-peak, \SI{7.5}{s} mean period) and a quadcopter UAV with an onboard camera. Across 31 motion-capture and 20 vision-based trials, the UAV landed every time, with median times to contact of \SI{6.5}{} and \SI{7.3}{s}; 77\% and 50\% landed on the first attempt, with mean deck-relative speeds of \SI{0.29}{} and \SI{0.27}{m/s}, respectively, at the instant of thrust cutoff, which is the last speed under the policy's control. Supplementary video: \url{https://youtu.be/S5hDkrSZJt4}

\end{abstract}


\section{Introduction}
\label{sec:intro}
Autonomous landing of a vertical take-off and landing (VTOL) capable unmanned aerial vehicle (UAV) on a moving ship deck matters for maritime logistics and naval recovery. The vehicle must track a deck that translates and rotates, choose when to descend, and touch down within tight limits on relative velocity, attitude, and deck size. For ship-deck motions such as roll (about $X$), pitch (about $Y$), sway (along $Y$), and surge (along $X$), the vehicle can null the relative translation it currently observes, and it can decide to stop or continue the descent according to the tilt it observes at that moment. That observation is a sufficient statistic for landing. Early deck-landing systems accordingly focused on recovering the relative deck pose and controlling the vehicle's angular and horizontal motion. Monocular and marker-based methods established this pipeline on laboratory motion platforms, simulated decks, and wave-perturbed small surface vessels~\cite{Sanchez_Lopez_2013,Polvara_2018,Shastry,shankarVFS}. 
These systems established the perception and feedback required for landing under observable angular and lateral deck motion. With vertical deck travel absent, the current relative pose remained sufficient for deciding the control response.

Horizontal motion is a tracking problem; heave (along $Z$), on the other hand, is a timing problem because a poorly timed descent can bring the vehicle into contact with a rising or upward-accelerating deck at a high relative velocity. The quantity that determines contact is now the relative velocity at an instant that has not yet arrived, so the current relative pose no longer determines the correct action. A constant descent makes the difficulty concrete. Taking every sample of a measured ship-heave record~\cite{sconedataset} as an instant at which the vehicle could arrive, and descending at a constant \SI{0.5}{m/s}, the rate an off-the-shelf autopilot uses to land, the relative speed at contact exceeds \SI{0.5}{m/s} at half of them, because the deck is rising as often as it is falling. A landing attempt must therefore be timed against motion that has not yet occurred, and a descent must be abandoned once safe contact can no longer be achieved. The landing must also end by itself. An autopilot disarms only after the airframe stops moving, and a heaving deck keeps it moving, so the vehicle would sit on the deck with its motors still running and never disarm. The policy therefore cuts the thrust itself, once it is within \SI{7}{cm} of the deck. Cutting at contact instead is possible, since the camera gives the relative height, but a vehicle that is still producing thrust when the gear meets a moving deck can slide or tip before it settles. Dropping the last few centimeters lets it arrive with enough weight on the gear to settle in place. This operational insight motivated the adoption of a proximity-based disarm logic. The policy steers the descent up to that instant, and the vehicle falls unpowered after it, so the speed the policy achieves and the speed at contact are two different numbers.

Classical approaches decompose this task into estimation, motion prediction, descent planning, and safety logic. Precision landing on a vertically oscillating platform has been achieved by tracking its height~\cite{Nguyen2020HeavingLanding} and by coupling a UKF estimate of known multi-sinusoidal motion to time-optimal trajectory generation~\cite{Hu2018TimeOptimalLanding}. Such pipelines rest on a forecast of the ship's motion, from Kalman estimators~\cite{Sidar1975CarrierMotion,Triantafyllou1983ShipMotion,Richter2014HeaveEstimation} to learned time-series models~\cite{Keijzer2022ShipHeave,Geng2023EMDPSOLSTM,Yang2025ReservoirComputing}, coupled to a planned descent~\cite{scaled_experiments}. In maritime experiments, an FFT--Kalman observer and MPC reached the deck near zero roll and pitch with a forecast of about \SI{1}{s}, treating vertical motion as negligible and scoring success on tilt alone~\cite{Gupta2023HarshWaters}; a later system from the same group added onboard vision and landed on a rocking USV, reporting its hardware trials as roll and pitch~\cite{Prochazka2024AgileLanding}. Distributed MPC has selected a calm patch of water and a low-tilt touchdown time from a known wave model, in a two-dimensional simulation with a communicating vessel~\cite{stephenson2024distributed}. Prediction-driven implementations still require an explicit forecast, a hand-designed phase structure, and a tuned descent trigger.


Reinforcement learning (RL) offers a route to coordinating the coupled tracking and timing objectives within a policy. Landing policies have progressed from visual touchdown and inclined surfaces to continuous tracking of horizontally moving pads~\cite{Polvara2018EndToEnd,RodriguezRamos2019MovingPlatform,Xie2020TrackingLanding,Kooi2021Inclined,Goldschmid2024MovingPlatforms}. End-to-end image-based landing with learned three-axis velocity commands has been demonstrated on a ground vehicle following a horizontal path at up to \SI{1}{m/s}; the study identifies memory as a requirement for more complex platform motion~\cite{ladosz2024autonomous}. Robust policies have also been trained for a sub-scale ship and wind disturbances~\cite{saj2025robust}. That policy commands roll and pitch while a separate command executes the vertical landing, so heave stays outside the action space and the descent is not decided against deck motion.

The closest learned heave systems broaden the motion and validation envelope. A policy that infers platform motion from two successive sparse visual observations and commands attitude and thrust was evaluated in 411 onboard trials on a Stewart platform and a wave-excited pool float~\cite{angelis2026visionbased}. The success criterion of Angelis et al.~\cite{angelis2026visionbased} allows \SI{1.5}{m/s} of relative speed at contact. The two test platforms of Angelis et al. each exhibit one half of the problem: the Stewart platform moves \SI{0.25}{m} but follows a single sine wave, and the pool float moves irregularly but only \SI{0.13}{m}. The algorithm in our work does both at once, following recorded ship motion over \SI{0.70}{m} peak to peak, and a touchdown is scored against a \SI{0.5}{m/s} bound on the relative speed. WaveLander~\cite{li2026wavelander} starts from the same premise as our work, learning when to descend, hold position, and retreat through a recurrent policy that commands a scalar vertical-velocity reference. The WaveLander policy is trained without privileged information; the policy in our work gives the critic future deck height that the actor never observes, so the descent decision is learned against motion that the policy itself cannot see. The WaveLander training model constrains the vehicle to one vertical prismatic joint and drives the deck with synthetic multi-sinusoidal motion plus correlated noise; our work considers ship-heave motion~\cite{sconedataset}, which is not purely periodic in nature. In WaveLander, lateral tracking is a fixed controller; success is judged on attitude, and one deployment test is reported. In our work, the policy commands the lateral channels itself and is flown 51 times on a physically heaving deck emulator platform.

The literature leaves three gaps. First, heave timing is seldom learned from instantaneous measurements of ship motion. Second, policies are trained above a simulator-side controller that only approximates the autopilot that will fly them; putting the real autopilot in the loop fixes the mismatch, but runs one vehicle at a time~\cite{jacinto2024pegasus}, which rules out the parallel training that these policies need. Third, learned maritime landing algorithms reach deployment without a runtime safety check on the descent they commit to~\cite{saj2025robust,angelis2026visionbased,li2026wavelander}.

This paper closes all three gaps. The deck position a few seconds ahead decides the landing. A vehicle in flight cannot know it, but the simulator does, because it replays a recorded sea state. Only the critic is given that future deck height, and only during training. The actor is recurrent, so it reads the state of the deck from what it has just seen and decides when to descend. No forecast window or peak-selection rule is prescribed. A feed-forward actor sees a single observation and cannot do this, so the privileged critic has nothing to teach it; feed-forward runs did not learn to land, and the two are not compared here. For the second gap, the policy is trained above a Torch implementation of the PX4 cascade~\cite{meier2015px4}, and it commands world-frame velocity and yaw rate. The deployed autopilot takes those same four channels, so the interface does not change between simulation and flight.

For the third gap, a control-barrier-function (CBF)~\cite{ames2017cbf} filter wraps the deployed policy in an outer loop and checks the vertical velocity it commands at every tick. The check reads the deck-relative vertical state and passes only while a feasible stopping margin remains. Once no margin remains, the filter overrides the command and aborts to hover, so a descent that has run out of stopping margin becomes a retry.

To the best of our knowledge, this is the first learned ship-deck landing policy trained with privileged future deck motion and checked at run time by a control-barrier-function monitor on flight hardware.
The barrier is used in its standard role, and the work makes no claim of a new safe-learning theory. This work combines these elements and demonstrates their use for landing a vehicle on a physically heaving deck.

\begin{figure}[!t]
    \centering
    \begin{minipage}[b]{0.50\linewidth}\centering
        \includegraphics[clip,height=2.25cm,width=\linewidth,keepaspectratio]{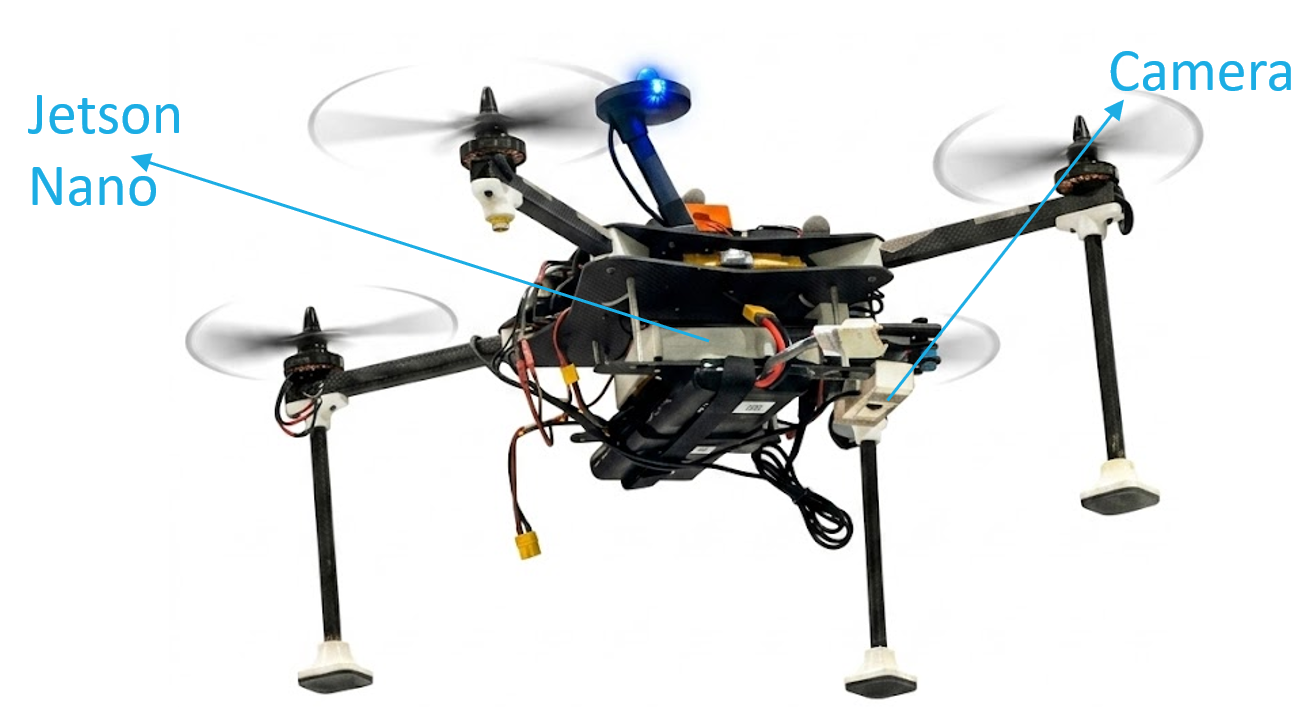}\\[2pt]
        {\footnotesize (a)}
    \end{minipage}\hfill
    \begin{minipage}[b]{0.49\linewidth}\centering
        \includegraphics[clip,height=2.25cm,width=\linewidth,keepaspectratio]{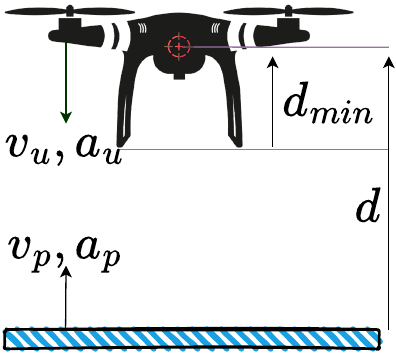}\\[2pt]
        {\footnotesize (b)}
    \end{minipage}
    \caption{(a) Quadrotor used for the heave-landing experiments. (b) Deck-relative vertical quantities used by the CBF filter.}
    \label{fig:teaser}
\end{figure}

The main contributions of this work are as follows:
\begin{enumerate}
    \item a recurrent landing policy for a heaving deck, trained with an asymmetric actor--critic, where the critic sees the next \SI{6}{s} of deck motion and the actor only the deck-relative state available onboard, and which commands the autopilot through its own velocity interface;
    \item a control-barrier-function filter around the deployed policy that aborts and retries when the stopping margin is exhausted, with the same margin used as a training reward, which halves the median simulated contact speed against the same policy trained without it;
    \item 51 hardware landings on a deck replaying measured ship heave at \SI{0.70}{m} peak to peak, 31 under motion capture and 20 under onboard vision, all successful, with 97\,\% and 95\,\% of disarms inside \SI{0.5}{m/s}; the disarm speed the policy controls and the contact speed the airframe absorbs are reported separately, and the unpowered fall between them is measured.
\end{enumerate}
The rest of the paper is organized as follows. Sec.~\ref{sec:heave} defines the landing task and describes the policy's observation, action, architecture, and training. Sec.~\ref{sec:cbf} develops the barrier on the deck-relative vertical state and the runtime check and retry it drives. Sec.~\ref{sec:setup} describes the vehicle, the deck emulator, and the criteria by which a landing is judged. Sec .~\ref {sec:results} presents the simulation benchmark of five controllers and the hardware study of 51 landings under motion capture and onboard vision. Sec.~\ref{sec:conclusion} concludes and sets out the future work.

\section{Heave-Landing Task and Policy}
\label{sec:heave}
\subsection{Task Definition}
With world-frame states $\mathbf{p}_{u},\mathbf{v}_{u}$ for the UAV and $\mathbf{p}_{p},\mathbf{v}_{p}$ for the platform, the deck-relative state is $\mathbf{p}_{\mathrm{rel}}=\mathbf{p}_{u}-\mathbf{p}_{p}$ and $\mathbf{v}_{\mathrm{rel}}=\mathbf{v}_{u}-\mathbf{v}_{p}$; the platform translates only in heave. Each episode starts with the UAV above the deck, offset horizontally and slightly perturbed. It must align with the deck center, hold a level attitude, choose when to descend, and reach first contact within the horizon at a low deck-relative vertical speed. No future deck motion is available at deployment, so the descent decision comes from the observation history alone, making the heave landing a partially observable finite-horizon problem. Touchdown and failure conditions follow the training setup. Fig.~\ref{fig:system} summarizes the task, the training loop, and the deployment path.

\subsection{Observation, Action, and Policy Architecture}
\label{sec:priv}
At \SI{20}{Hz}, the actor receives only quantities available from the onboard estimator and deck-relative pose pipeline:
\begin{equation}
    \mathbf{o} = \big[\mathbf{p}_{\mathrm{rel}},\; \mathbf{v}_{\mathrm{rel}},\; \mathbf{q}_{\mathrm{rel}},\; \mathbf{g}_{b},\; \mathbf{u}_{t-1}\big]\in\mathbb{R}^{17},
    \label{eq:obs}
\end{equation}
Here $\mathbf{p}_{\mathrm{rel}}$, $\mathbf{v}_{\mathrm{rel}}$ and $\mathbf{q}_{\mathrm{rel}}$ are the vehicle's position (3), linear velocity (3) and attitude quaternion (4) in the deck frame, $\mathbf{g}_{b}$ is gravity in the body frame (3), and $\mathbf{u}_{t-1}$ is the previous command (4) in the units of eq.~\eqref{eq:contract}. None of the 17 channels depends on a world-frame origin. The actor maps the observation history to the world-frame velocity and yaw-rate command
\begin{equation}
    \mathbf{u}_t = [v_x,\; v_y,\; v_z,\; \dot{\psi}]^T,
    \label{eq:contract}
\end{equation}
with $|v_{x,y}|\le\SI{0.8}{m/s}$, $v_z\in[-0.8,\,1.0]$~m/s and $|\dot\psi|\le 35^{\circ}$/s. This velocity-setpoint interface uses the flight controller's estimator and inner loops and remains unchanged between simulation and flight.

A single observation in eq.~\eqref{eq:obs} does not determine the deck's future trajectory, so the actor is recurrent: two stacked Gated Recurrent Units (GRU) layers~\cite{cho2014gru} of sizes $[128,64]$ feed a 32-unit Exponential Linear Unit (ELU) multilayer perceptron (MLP) and a four-dimensional action head. Its memory encodes the latent heave phase from the observation history.

Training uses an asymmetric actor--critic~\cite{pinto2018asymmetric}. Besides the 17 actor channels, the critic receives the deck-relative angular velocity (3), the next \SI{6}{s} of deck height at \SI{20}{Hz} (120), and the exact world-frame velocities of the UAV and deck. Its 146-dimensional input passes through a stacked GRU of sizes $[512,256]$, a 128-unit ELU MLP, and a scalar value head. The critic therefore values a descent against the motion that follows, and recurrence lets the actor tie that signal to its own observation history. 

\subsection{Training Setup}
\label{sec:training}
\paragraph*{Recorded deck motion and data split}
The environment replays five \SI{1800}{s} heave records (D1H1--D1H5, abbreviated D1--D5) from the SCONE dataset~\cite{sconedataset} at \SI{20}{Hz}. D1, D2, D4, and D5 are used for training; D3 is used only for validation and testing. Every reset samples a training record, a start index, and a deck-height bias in $[0.5,2.5]$~m. Each record is centered and normalized to $[-1,1]$, so scaling by $0.35$ gives \SI{0.70}{m} commanded peak-to-peak heave.




\paragraph*{Simulation and action execution}
Episodes reset with the UAV at \SI{3.5}{-}\SI{5.0}{m} and $x,y\in[-1,1]$~m about the deck, with small attitude and velocity perturbations, and the platform moves in every environment. Domain randomization covers mass, action delay, the position, velocity, attitude, and projected-gravity estimates, per-rotor thrust scale, motor lag, and velocity-loop gains. A batched Torch re-implementation of a PX4-like multicopter cascade executes each velocity setpoint rather than applying it as a force, running with the physics at \SI{200}{Hz} beneath the \SI{20}{Hz} policy, so training and deployment share the interface of eq.~\eqref{eq:contract}.

\paragraph*{Policy optimization}
Training uses PPO~\cite{schulman2017ppo} in RSL-RL with 4096 parallel Isaac Lab environments, 48 steps per environment and update, six mini-batches, five epochs, clip 0.2, $\gamma=0.99$ and $\lambda=0.95$. Training from scratch takes about 3 h on a desktop with an NVIDIA RTX 5090, at which point the good-touchdown rate has converged. The same weights are then fine-tuned in a 16-environment run, with the deployed vision pipeline in the loop, so the actor acts on the filtered estimates it receives in real time. 

\subsection{Reward and Episode Termination}
\label{sec:reward}
At each policy step $r_t=\Delta t\sum_i w_i\phi_i$ with $\Delta t=\SI{0.05}{s}$. Dense terms center the UAV over the deck and penalize tilt, yaw error, and aggressive commands; they regularize flight and are not specific to heave. Two terms set the landing behavior. The first charges violations of the barrier margin $h$ of Sec.~\ref{sec:cbf}, as a Huber penalty on $\max(0,-h)/\SI{0.25}{m}$, which pushes the policy towards descents that keep a stopping margin; Sec.~\ref{sec:cbfrl} isolates its effect. The second is a sparse bonus at first contact $T$, when the body contact force first exceeds \SI{2}{N}. With $s=\max(-v_{\mathrm{rel},z}^{\mathrm{pre}},0)$ the pre-impact closing speed, $\rho=\|\mathbf p_{\mathrm{rel},xy}\|_2$ and $\bar\psi=\operatorname{wrap}(\psi)$, a good touchdown is
\begin{equation}
\begin{aligned}
G=T\land\{&s\le\SI{0.25}{m/s},\; \rho\le\SI{0.20}{m},\\
&|\phi|,|\theta|\le12^\circ,\; |\bar\psi|\le15^\circ\},
\end{aligned}
\label{eq:good}
\end{equation}
scored higher the further inside those bounds the vehicle lands, and the slower the deck moves. The \SI{0.25}{m/s} in eq.~\eqref{eq:good} is the simulation limit; in flight the bound is \SI{0.5}{m/s}. Terminations other than touchdown are heavily penalized. Hardware trials use the disarm criterion of Sec.~\ref{sec:setup}.

\section{CBF Safety Filter on the Learned Policy}
\label{sec:cbf}
A learned policy can enter a descent from which it can no longer slow to the touchdown speed in the clearance that is left. The deployed actor is therefore wrapped in a control-barrier-function filter~\cite{ames2017cbf,ames2019cbf} on the deck-relative vertical state. Let $d=z_u-z_p$ be the separation between the vehicle center of gravity and the deck pose origin (marker frame), $d_{\min}$ the required clearance (absorbing landing-gear length and the origin-to-surface offset), and $v_r=v_u-v_p$ the relative vertical velocity (Fig.~\ref{fig:teaser}(b)), with $\dot d=v_r$ and $\dot v_r=a_u-a_p$ for vehicle and deck vertical accelerations $a_u$, $a_p$. The policy commands a velocity setpoint $v_z$ in eq.~\eqref{eq:contract}; the filter does not model the autopilot's velocity loop but treats its response $a_u$ as the input and reads it from the onboard inertial measurement, so the barrier is checked against the acceleration the vehicle is actually producing. With $v_L$ the admissible touchdown speed and $a_{\mathrm{rel,max}}$ a conservative bound on relative braking acceleration, the clearance needed to slow the closing speed to $v_L$ is
\begin{equation}
    d_{\mathrm{stop}}(v_r) = \frac{(v_r^{-})^2 - v_L^2}{2\,a_{\mathrm{rel,max}}},\qquad
    v_r^{-} = \tfrac{1}{2}\big(v_r - \sqrt{v_r^2+\varepsilon}\big),
    \label{eq:dstop}
\end{equation}
where $v_r^{-}\le 0$ is a smooth approximation of $\min(v_r,0)$; $d_{\mathrm{stop}}$ is negative once the closing speed is below $v_L$, since contact at up to $v_L$ needs no braking clearance. The barrier and safe set are
\begin{equation}
    h(d,v_r) = (d - d_{\min}) - d_{\mathrm{stop}}(v_r),\qquad \mathcal{C}=\{h\ge 0\}.
    \label{eq:h}
\end{equation}
Differentiating along the relative dynamics, with $\sigma(v_r)=\partial v_r^{-}/\partial v_r=\tfrac12\big(1-v_r/\sqrt{v_r^2+\varepsilon}\big)\in(0,1)$, gives a form that is control affine in $a_u$,
\begin{equation}
    \dot h = \underbrace{v_r - \beta(v_r)\,a_p}_{L_f h} + \underbrace{\beta(v_r)}_{L_g h}\,a_u,\qquad
    \beta = \frac{-\,v_r^{-}\,\sigma}{a_{\mathrm{rel,max}}}\ \ge 0 .
    \label{eq:hdot}
\end{equation}
$L_g h=\beta$ is strictly positive whenever the vehicle is closing and vanishes when it climbs, so braking harder than the deck always increases $h$ and the constraint is well posed throughout the descent. Forward invariance of $\mathcal{C}$ is enforced through the linear class-$\mathcal{K}$ condition
\begin{equation}
    L_f h + L_g h\,a_u + \gamma h \ge 0
    \;\Longleftrightarrow\;
    a_u \ge a_p - \frac{v_r + \gamma h}{\beta(v_r)}\quad(\beta>0),
    \label{eq:cbfcond}
\end{equation}
which for a closing vehicle ($v_r^{-}\approx v_r$, $\sigma\approx1$) reads $a_u-a_p\ge a_{\mathrm{rel,max}}(1-\gamma h/|v_r|)$: no braking is demanded deep inside $\mathcal{C}$, and the full $a_{\mathrm{rel,max}}$ on its boundary. Rather than clipping $a_u$ as in the minimum-norm filter of~\cite{ames2017cbf}, eq.~\eqref{eq:cbfcond} is used as an abort test on the measured $a_u$: a descent needing more braking than the vehicle is producing has already mistimed the deck. The flight code evaluates eq.~\eqref{eq:hdot} with $\sigma\equiv1$ and $v_r^{-}$ in place of $v_r$ in the first term, which coincides with it while closing and is conservative while climbing.

\paragraph*{Runtime check and retry}
At each \SI{20}{Hz} tick the node evaluates $h$ and $\dot h+\gamma h$ from the deck-relative estimate of Sec.~\ref{sec:setup} and the accelerations, low-pass filtered at \SI{5}{Hz}: IMU and measured deck acceleration under motion capture, the finite-differenced relative velocity under vision. If $h\le0$ or $\dot h+\gamma h<0$ on three consecutive ticks, the node overrides all four channels, climbs to hover altitude, holds for \SI{4}{s}, then re-enables the policy with the previous-action input and GRU hidden state cleared, so the retry starts without memory of the aborted approach. Parameters, tuned empirically: $d_{\min}=\SI{0.263}{m}$, $\gamma=\SI{3.6}{s^{-1}}$, $v_L=\SI{0.3}{m/s}$, $a_{\mathrm{rel,max}}=\SI{0.7}{m/s^2}$, $\varepsilon=10^{-4}$.

\section{Experimental Setup}
\label{sec:setup}
\paragraph*{Hardware}
Experiments were performed using a 3-DoF parallel manipulator acting as a deck emulator platform, with \SI{\pm 35}{cm} of heave travel, replaying the scaled heave motion of the test set specified in Sec.~\ref{sec:training}. The aerial unmanned vehicle (Fig.~\ref{fig:teaser}(a)) is a \SI{3.2}{kg} custom assembled quadrotor with Cube Orange+ as a Flight Control Unit (FCU) running ArduCopter autopilot firmware, with a downward-facing Logitech C922 camera and an NVIDIA Jetson Orin Nano that runs the vision-estimator, GRU actor and CBF filter onboard; deck-relative pose arrives at \SI{20}{Hz} and the policy runs at \SI{20}{Hz}. One policy inference takes \SI{5.0 \pm 2.4}{ms} on the Orin Nano. Ground truth of states comes from a \SI{100}{Hz} Qualisys motion-capture system.
\paragraph*{State estimation}
Under vision, the deck-relative state is estimated onboard. The downward-facing camera detects a fractal ArUco marker~\cite{fractal_original} on the deck at \SI{20}{Hz}, and a Multiplicative Extended Kalman Filter (MEKF)~\cite{leishman2013multiplicative} fuses the detections into the estimate

\begin{equation}
    \hat{\mathbf{x}} = \big[\hat{\mathbf{p}}_{\mathrm{rel}},\; \hat{\mathbf{v}}_{\mathrm{rel}},\; \hat{\mathbf{q}}_{\mathrm{rel}}\big],
    \label{eq:xhat}
\end{equation}
The three parts of eq.~\eqref{eq:xhat}, the deck-relative position, velocity, and attitude quaternion, fill the first ten channels of the observation in eq.~\eqref{eq:obs}. The body-frame gravity term comes from the autopilot's attitude and the previous command from the node itself. The barrier of Sec.~\ref{sec:cbf} reads $d$ and $v_r$ from the same estimate. Under motion capture, $\hat{\mathbf{x}}$ is taken from the Qualisys system instead, so the policy and the filter receive the same quantities in both studies and only their source changes.

\paragraph*{Metrics}
A trial succeeds if the vehicle commits inside the deck footprint at a deck-relative vertical speed $\le\SI{0.5}{m/s}$, for the reason given in Sec.~\ref{sec:intro}; a disarm meeting this bound is called good. Two speeds describe each landing, both the deck-relative vertical velocity $v_{\mathrm{rel},z}$ read at a different instant. The disarm speed $|v_{\mathrm{rel},z}(t_{\mathrm{d}})|$ is read at $t_{\mathrm{d}}$, when the policy cuts thrust, which it does once the gear is within 7 cm of the deck. It is the last speed the policy sets, so the policy is judged on it. The contact speed $|v_{\mathrm{rel},z}(t_{\mathrm{TD}})|$ is read at first contact $t_{\mathrm{TD}}$ from \SI{100}{Hz} motion capture, \SI{0.1}{} to \SI{0.2}{s} later, with the vehicle unpowered in between. It is what the airframe absorbs. Both are reported.

\section{Results}
\label{sec:results}

\subsection{Benchmark and Ablation in Simulation}
\label{sec:cbfrl}

\begin{figure}[!t]
    \centering
    \includegraphics[width=0.99\linewidth]{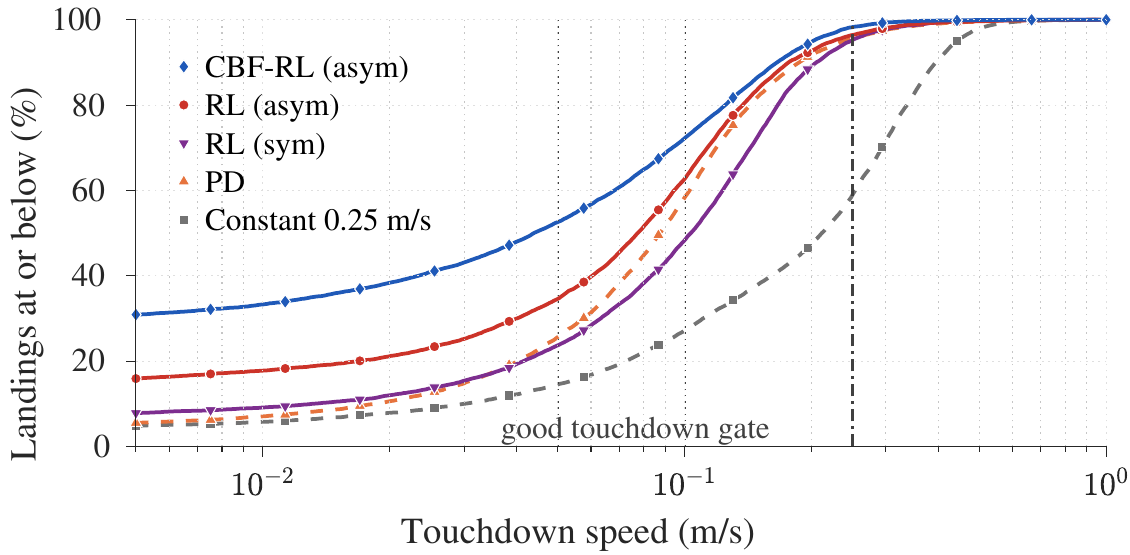}
    \caption{Simulation ablation study of various methodologies.}
    \label{fig:cbfrl}
\end{figure}

To benchmark the method before the hardware experiments, five controllers were compared at scale in the training simulator: 1024 environments, about 10\,500 touchdowns each, all observing the simulator's state with the estimate noise of Sec.~\ref{sec:training}. Fig.~\ref{fig:cbfrl} plots the share of landings at or below each touchdown speed, read one policy step before first contact so that braking during the collision cannot mask it. CBF-RL (asym) is the complete method; RL (asym) is trained identically without the barrier term of Sec.~\ref{sec:reward}; RL (sym)'s critic no longer receives the \SI{6}{s} deck forecast. Proportional Derivative (PD) tracks the deck-relative position and velocity and commits to a straight descent. Constant descends at \SI{0.25}{m/s}, the touchdown speed the simulation training reward aims for.

The barrier term shifts the whole distribution to lower speeds and halves the median, from \SI{0.077}{} to \SI{0.044}{m/s}. The deck forecast in the critic is what lifts the learned policies above the tuned controller: without it, the learned policy lands no better than PD. PD clears the \SI{0.25}{m/s} gate in most landings but lands gently only half as often as CBF-RL. The constant descent, commanded at the target speed itself, exceeds it in 41\,\% of landings, because the deck's own vertical speed adds to the commanded rate. The asymmetric CBF-RL curve lies above every other at every speed; it was therefore the policy taken to hardware.

\subsection{RL Landing on the Heaving Deck}
\begin{figure*}[!t]
    \centering
    \begin{subfigure}[t]{0.42\textwidth}
        \includegraphics[width=\linewidth]{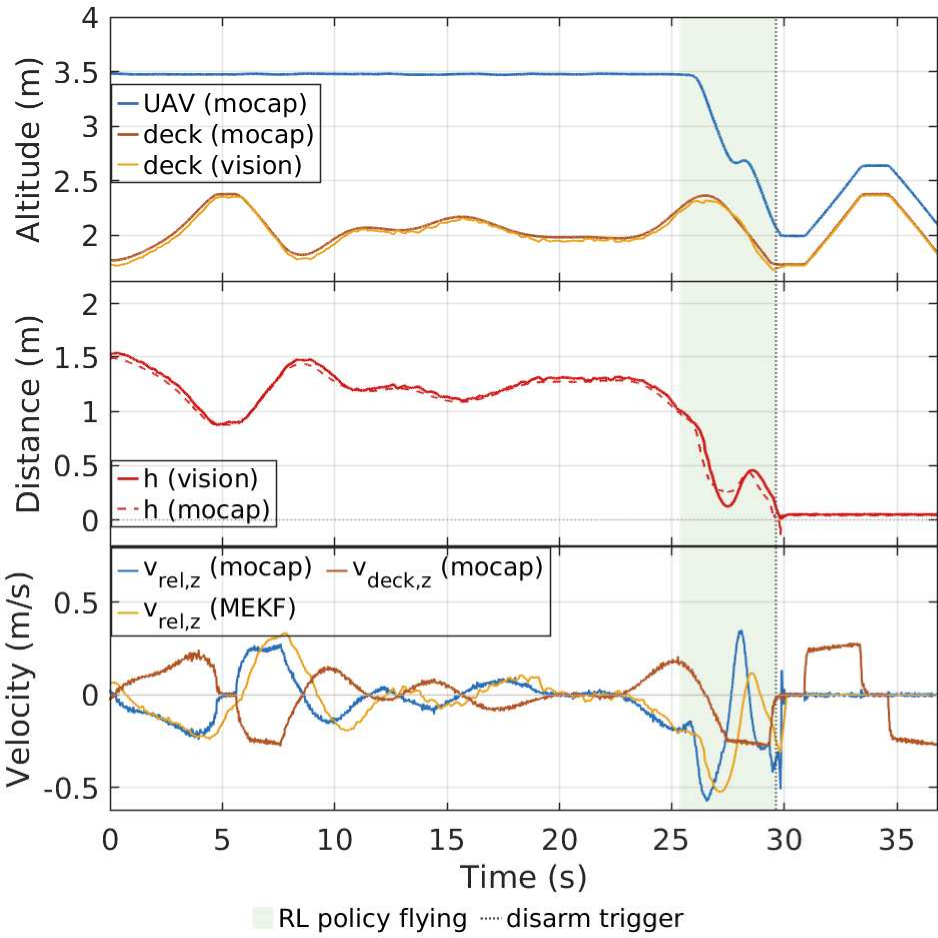}
        \caption{single descent}
        \label{fig:landing:oneshot}
    \end{subfigure}\hfill
    \begin{subfigure}[t]{0.42\textwidth}
        \includegraphics[width=\linewidth]{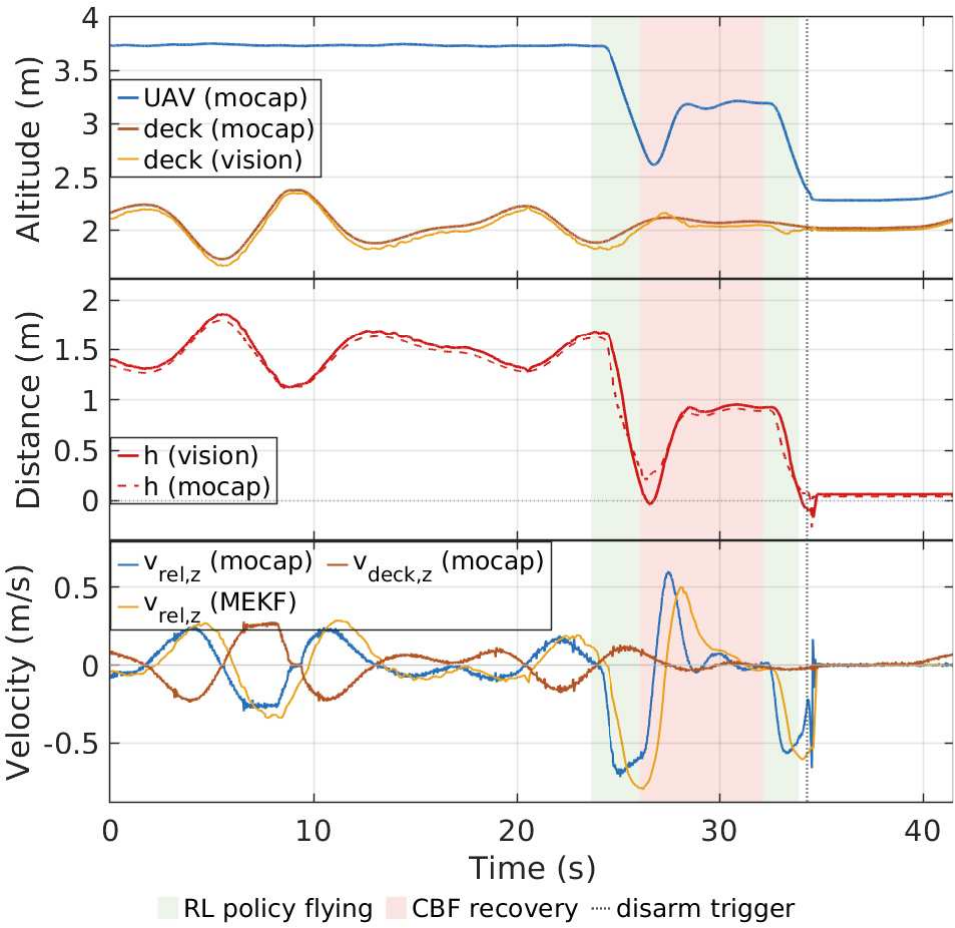}
        \caption{barrier-triggered retry}
        \label{fig:landing:recovery}
    \end{subfigure}
    \caption{Vision-based RL landings on the heaving deck.}
    \label{fig:landing}
\end{figure*}

\begin{figure}[!t]
    \centering
    \includegraphics[width=\linewidth]{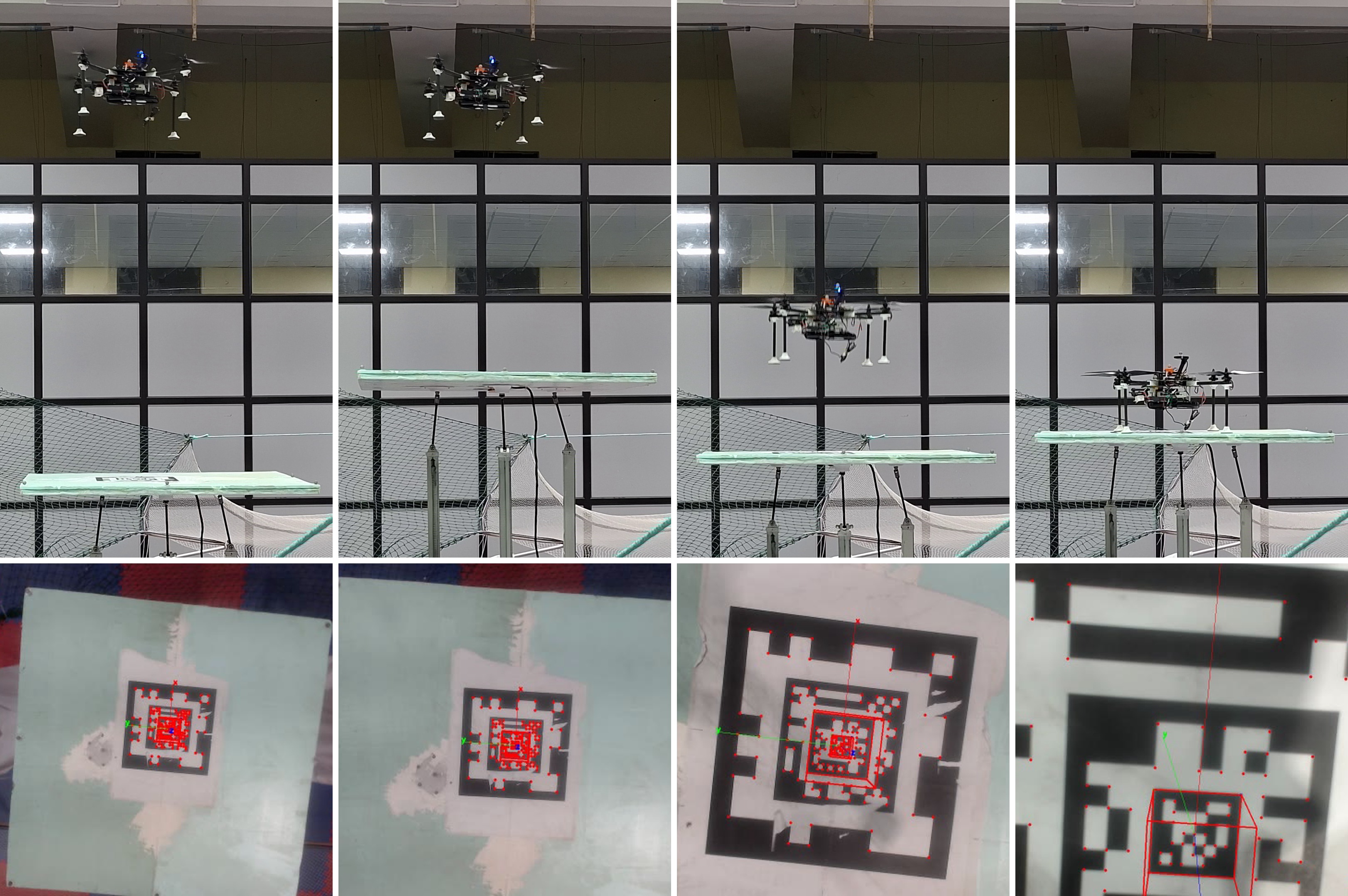}
    \caption{Landing sequence, hover to touchdown. Top: external view. Bottom: onboard camera with marker detections.}
    \label{fig:snaps}
\end{figure}

Fig.~\ref{fig:landing} shows two measured vision landings with the barrier active. In Fig.~\ref{fig:landing:oneshot} the policy descends once and lands; it disarms at a deck-relative speed of \SI{0.35}{m/s} and contacts at \SI{0.46}{m/s}, \SI{0.16}{s} later. In Fig.~\ref{fig:landing:recovery}, the barrier finds the stopping margin exhausted on the first descent, returns the vehicle to hover, and the policy lands on the next descent, disarming at \SI{0.26}{m/s} and contacting at \SI{0.58}{m/s} after \SI{0.22}{s}. The panels show the vehicle and deck altitudes, the barrier value $h$, and the deck-relative and deck vertical velocities, with the disarm trigger marked. Where the onboard estimate is drawn over motion capture, the two traces also show how closely the vision pipeline follows the true state through the descent. Fig.~\ref{fig:snaps} shows the corresponding touchdown sequence.

\begin{figure}[!t]
    \centering
    \includegraphics[width=\linewidth]{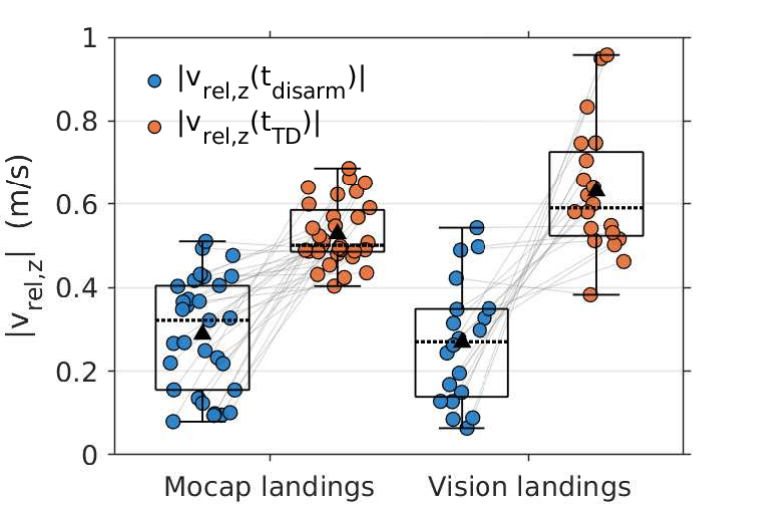}
    \caption{Deck-relative vertical velocity $|v_{\mathrm{rel},z}|$ at disarm
    $t_{\mathrm{d}}$ and at first contact $t_{\mathrm{TD}}$.}
    \label{fig:tdbox}
\end{figure}

\begin{figure}[!t]
    \centering
    \includegraphics[width=\linewidth]{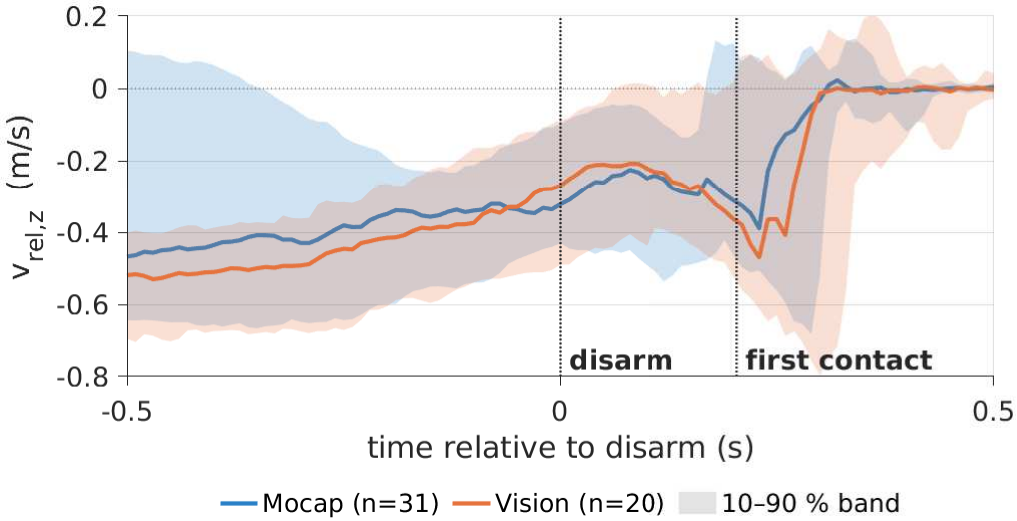}
    \caption{Deck-relative closing speed from \SI{0.5}{s} before to \SI{0.5}{s} after disarm, median and 10--90\,\% band over 31 motion-capture and 20 vision landings. The second dotted line marks the median first contact, \SI{0.20}{s} after disarm; landed runs settle to zero.}
    \label{fig:ensemble}
\end{figure}

\begin{figure}[!t]
    \centering
    \includegraphics[width=\linewidth]{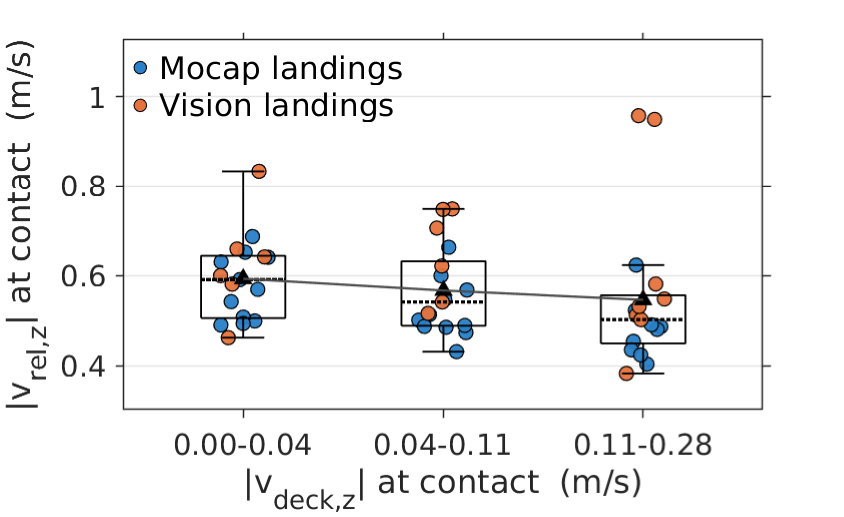}
    \caption{Deck-relative vertical speed at first contact, binned by deck speed at contact.}
    \label{fig:vsdeck}
\end{figure}

Fig.~\ref{fig:tdbox} reports the deck-relative speed at disarm and at first contact for both studies, in every case measured by motion capture. Every one of the 31 motion-capture and 20 vision estimated trials ended in a landing. Judged by the criterion of Sec.~\ref{sec:setup}, 30 of the 31 motion-capture
disarms and 19 of the 20 vision disarms are good (97\,\% and 95\,\%), and 48\,\%
and 60\,\% lie inside the barrier's \SI{0.3}{m/s}. 
At the trigger the closing speed is \SI{0.288}{m/s} ($\pm$\,0.135) under motion
capture and \SI{0.269}{m/s} ($\pm$\,0.145) under vision; by contact it has risen to
\SI{0.528}{m/s} ($\pm$\,0.076) and \SI{0.631}{m/s} ($\pm$\,0.153) respectively. The velocity increases after the disarm. Fig.~\ref{fig:ensemble} shows how the velocity changes with respect to time after the disarm command is sent. For about the first \SI{0.1}{s} the closing speed still falls, so the rotors are still braking the descent; the speed then grows into contact with a slope that steepens. The closing acceleration averages \SI{-0.3}{m/s^2} over the first half of the window and \SI{2.5}{m/s^2} over the second, about a quarter of gravity, so the thrust holds briefly after the command and is only part-way through its decay when the UAV touches the deck. The response window lasts
\SI{0.190}{s} ($\pm$\,0.064) for mocap landings and \SI{0.223}{s} ($\pm$\,0.081) for vision landings, and the vehicle drops
\SI{0.058}{m} ($\pm$\,0.007) and \SI{0.072}{m} ($\pm$\,0.017) respectively through it with the rotors
spinning down. The deck's own speed changes by \SI{0.02}{m/s} over so short an
interval, so this fall distance sets the contact speed. The policy has authority
only up to $t_{\mathrm{d}}$, which is the cost of cutting thrust autonomously, and it
is why the policy is judged at the trigger and contact is reported as the physical
outcome. The three vision landings near \SI{0.9}{m/s} in Fig.~\ref{fig:tdbox} show
this directly: each disarmed below \SI{0.1}{m/s}, then ran a window of \SI{0.31}{}
to \SI{0.36}{s} and fell \SI{0.08}{} to \SI{0.11}{m} before contact. 

Deck motion does not degrade the contact itself (Fig.~\ref{fig:vsdeck}). Landings are binned by the deck speed at the instant of contact, and the contact speed does not rise with it. The deck speed at contact sits at the 48th percentile of the deck's own speed distribution over the preceding approach, so the policy does not hold out for a slow phase of the heave. It tracks continuously and contacts wherever it arrives.


Table~\ref{tab:attempts} reports how many descents each landing needed. A retry is counted when the barrier finds the descent unsafe, and the vehicle returns to hover before descending again.
Under motion capture, 77\% of landings reached the deck on the first descent, and under vision, 50\%. 
From the start of the landing procedure to first contact, the policy took a median of \SI{6.5}{s} under motion capture and \SI{7.3}{s} under vision. The policy commands from its first tick and needs no observation record before it can descend. Fig.~\ref{fig:landing:recovery} shows one such
retry, with the barrier returning the vehicle to hover and the second descent
reaching the deck. 
Every run in both studies ended in a landing, and retrying did not degrade the touchdown. Under motion capture, the landings that retried contacted the deck at a mean of \SI{0.507}{m/s} and those which did not, at \SI{0.558}{m/s}, a difference well inside the run-to-run spread.

\begin{table}[!t]
    \centering
    \caption{Number of descents per landing.}
    \label{tab:attempts}
    \footnotesize\setlength{\tabcolsep}{3pt}
    \begin{tabular}{lccccc}
        \toprule
        Sensing & Attempt 1 & 2 & 3 & 4 & 5 \\
        \midrule
        Mocap, $n{=}31$  & 24 (77\%) & 0 & 3 (10\%) & 2 (6\%) & 2 (6\%) \\
        Vision, $n{=}20$ & 10 (50\%) & 10 (50\%) & 0 & 0 & 0 \\
        \bottomrule
    \end{tabular}
\end{table}

\paragraph*{Failure modes}
To assess the sensing available to both the policy and the barrier, the onboard estimates from the 20 vision trials were compared with motion capture. The position errors (estimate minus motion capture) were $-0.06\pm0.12$~m, $0.04\pm0.10$~m and $0.07\pm0.03$~m in $x$, $y$ and $z$, respectively; the corresponding velocity errors were $-0.02\pm0.11$ and $0.01\pm0.11$~m/s horizontally and $0.15\pm0.22$~m/s vertically. The vision-based vertical-velocity estimate lagged motion capture by about \SI{0.49}{s}. Shifting the vision trace earlier in time, as an offline check, leaves a residual error of $-0.01\pm0.07$~m/s and removes 91\,\% of the variance, confirming that the vertical error is delay rather than noise. Thus, at the cutoff trigger, the onboard estimator still reported the closing speed from approximately half a second earlier: across vision runs, it averaged \SI{0.53}{m/s} against a simultaneous motion-capture mean of \SI{0.27}{m/s}.

A landing is scored as a failure when thrust is cut at $|v_{\mathrm{rel},z}| > \SI{0.5}{m/s}$. Although all 51 trials reached the deck, only 49 satisfied this cutoff-speed criterion; the two disarms above \SI{0.5}{m/s} were \SI{0.510}{m/s} under motion capture, just over the bound, and \SI{0.543}{m/s} in a vision run whose velocity estimate had failed and read zero through the descent. A barrier-initiated return to hover is not considered a failure but is recorded separately as a retry. 

The delayed velocity also explains the higher retry rate under vision.  In ten of the vision landings, the barrier intervened once during the descent, returned the vehicle to hover, and the policy then landed on a second descent. For each intervention, the barrier was re-evaluated with the motion-capture state at the same instant. The true margin was still positive every time, between \SI{0.06}{} and \SI{0.37}{m}, so every intervention came early. In two of the ten, the true margin went negative within the next \SI{0.6}{s}, so the intervention anticipated a violation the deck was about to produce; in the other eight, it never did, and the intervention was caused by the lagged velocity estimate alone. A retry costs a few seconds, and no landing was lost, so the filter erred on the safe side. The barrier's guarantee holds for the state it is given: with exact relative height and velocity and braking within $a_{\mathrm{rel,max}}$, the class-$\mathcal{K}$ condition keeps $h \ge 0$. In flight, its inputs arrive late or fail, so a filter for deployment must also account for the age and validity of its estimate; that is left to future work. 

Vision-in-the-loop fine-tuning did little to close the sensing gap. Rendering the camera and running the detector limited that stage to 16 environments, and raising the count cut training throughput sharply, so the policy saw far fewer vision samples than state samples. Future work will move the whole pipeline into the loop, either through feature tracking that yields the relative pose directly, as in~\cite{angelis2026visionbased}, or by parallelizing the marker detector across environments.

\section{Conclusion and Future Work}\label{sec:conclusion}
This research proposed a method for ship-deck landing of a rotorcraft on a heaving deck with an asymmetric actor-critic whose critic sees the future deck and whose actor does not, and verified the learned descent through real-time hardware deployment with a stopping-distance barrier. On a deck emulator replaying measured ship heave at \SI{0.70}{m} peak to peak, all 51 trials landed. Under motion capture, 30 of 31 cut thrust within \SI{0.5}{m/s} of the deck, 77\,\% of them on the first descent. Under onboard vision, 19 of 20 did so, half of them on the first descent, the retries coming from a delayed velocity estimate rather than from the policy. 

Four directions follow from this work. The emulator reproduces deck kinematics alone, so future work will add air-wake and wind over the deck, and will replay scaled six-degree-of-freedom sea states rather than heave in isolation. Deck travel bounds the motion that can be flown indoors, and taking the vision-estimated pipeline outdoors removes that bound and the cooperative marker together. The barrier covers the vertical channel under an assumed braking bound, and extending it to the lateral channels and to a bound identified in flight would widen the guarantee. Sec.~\ref{sec:results} shows that the sensing, rather than policy, limits what the barrier can guarantee, so the estimator improvements described there are the first of these to pursue.


\bibliographystyle{IEEEtran}
\bibliography{references}

\end{document}